\documentclass[]{hdsr}

\graphicspath{{figs/}}

\begin{document}

% Larger bottom margin for the first page
\newgeometry{bottom=1.5in}

% Editorial staff will replace the following values:
% 1. Volume number
% 2. Issue number
% 3. Article DOI
% e.g. for Volume 2, Issue 3, DOI 12.345:
% \volumeheader{2}{3}{12.345}
% remove comment below for HDSR compile
%\volumeheader{0}{0}{00.000}

\begin{center}

  \title{Noisy Measurements Are Important, the Design of Census Products Is Much More Important}
  \maketitle

  % Start page numbering on second page. Must appear *after* \maketitle
  \thispagestyle{empty}
  
  \vspace*{.2in}

  % Authors and Affiliations
  \begin{tabular}{cc}
    John M. Abowd\upstairs{\affilone,*}
   \\[0.25ex]
   {\small \upstairs{\affilone} Cornell University} \\
%   {\small \upstairs{\affiltwo} Affiliation Two} \\
%   {\small \upstairs{\affilthree} Affiliation Three} \\
  \end{tabular}
  
  % Replace with corresponding author email address
  \emails{
    \upstairs{*}john.abowd@cornell.edu 
    }
  \vspace*{0.4in}

\begin{abstract}
\citet{mccartan:etal:2023} call for ``making differential privacy work for census data users.'' This commentary explains why the 2020 Census Noisy Measurement Files (NMFs) are not the best focus for that plea. The August 2021 letter from 62 prominent researchers asking for production of the direct output of the differential privacy system deployed for the 2020 Census signaled the engagement of the scholarly community in the design of decennial census data products. NMFs, the raw statistics produced by the 2020 Census Disclosure Avoidance System before any post-processing, are one component of that design---the query strategy output. The more important component is the query workload output---the statistics released to the public. Optimizing the query workload---the Redistricting Data (P.L. 94-171) Summary File, specifically---could allow the privacy-loss budget to be more effectively managed. There could be fewer noisy measurements, no post-processing bias, and direct estimates of the uncertainty from disclosure avoidance for each published statistic.
\end{abstract}
\end{center}

\vspace*{0.15in}
\hspace{10pt}

%remove comment below for HDSR compile
%\copyrightnotice

%remove comment on media summary for HDSR compile
%\section*{Media Summary}
%Only three uses of the U.S. Decennial Census of Population and Housing have direct constitutional and federal statutory foundations: (1) apportionment of the House of Representatives, (2) statistical support for redistricting every legislative body in the country, and (3) statistical support for the Population Estimates Program. These uses dominate the publication format and accuracy assessments of modern U.S. censuses. While academics like McCartan, Simko and Imai properly focus on valid statistical inferences based on published data, that is only one feature required to assess their fitness for use. The redistricting data would greatly benefit from research that modified the publication format so that the data could be produced with fewer constraints while remaining fit for their intended use.

\section{The Origins of the Noisy Measurement Files}
\label{sec:origins}

There are many, many uses of data from the U.S. Decennial Census of Population and Housing, but only three have direct constitutional and federal statutory foundations. The first, specified in Article 1 § 2 of the U.S. Constitution, is apportionment of the House of Representatives. The second, specified in the 1954 Census Act as amended (13 U.S. Code § 141), is statistical support for redistricting every legislative body in the country. The third is statistical support for the Census Bureau's Population Estimates Program, which is also authorized in the Census Act as amended (13 U.S. Code § 181). These uses dominate the publication format and accuracy assessments of modern U.S. censuses. While academics like McCartan, Simko and Imai properly focus on valid statistical inferences based on published data, that is only one feature required to assess their fitness for use. For the redistricting data considered here, the dominant use case is the ability to support accurate, approximately equal-population new voting districts, whose boundaries obviously cannot be specified in advance except for the political perimeter encompassing all districts of a particular legislature, that meet the requirements of the 1965 Voting Rights Act. This commentary suggests how research using the 2020 Census Noisy Measurement Files (NMFs) can inform the design of future decennial census data products in a meaningful way. 

The redistricting data NMF, released on June 15, 2023, and the massive demographic and housing characteristics NMF, released on October 23, 2023, are the first data publications by any statistical agency in the world of the raw output of a confidentiality protection system. They are, effectively, the harbinger of $21^{st}$ century replacements for public-use microdata files because they contain massively more information than is embodied in the official tabular releases. In particular, they contain information on every high-order interaction consistent with the publication schema of every variable in any published tabulation for a given population (persons or housing units) at every level of geography specified in the hierarchy used to create the tabular publications. That's a dense sentence, and it takes a while to sink in. Some details clarify what I mean.

The official 2020 Redistricting Data (P.L. 94-171) Summary File (redistricting data, hereafter) contains two tables of race and ethnicity counts for all persons, two tables of race and ethnicity counts for all adults, one table of population counts in households and major group quarters categories, and one table of counts of occupied and vacant housing units---approximately 1.5 billion linearly independent statistics. But the redistricting NMF also contains data for the race and ethnicity of all persons and all adults living in each major group quarters type. And these data are available for the same geographic hierarchy that was used to produce the official data---approximately 16 billion linearly independent statistics. The Census Bureau's willingness to consider releasing the NMFs, and its active solicitation of user input for that process began at the December 2019 Committee on National Statistics (CNSTAT) workshop on the 2020 Census Disclosure Avoidance System (DAS), where one of the authors of the DAS, William Sexton, reported:

\vspace{0.1in}
\noindent\emph{The useful point here is that the noisy measurements are (as promised by differential privacy) future-proof to any subsequent attacks, meaning that providing direct access to the noisy measurements need not require the use of the Census Bureau’s Research Data Centers as a broker. They could be released as an alternative product to the census. (However, the approach would require the Census Bureau to support alternative releases surrounding the decennial census.) Given resource constraints, Sexton said that the Census Bureau is awaiting feedback from the user community before committing to producing alternative data releases and products.} \citep[p. 155]{NASEM:2020}
\vspace{0.1in}

Nothing in the August 2021 letter to the Census Bureau from 62 prominent researchers \citep{dwork:etal:2021} that further alerted the academic community to the role of the NMFs in the 2020 DAS foreshadowed the enormous volume of information in those noisy measurements that was not part of the official data. The official 2020 Demographic and Housing Characteristics File (DHC) contains approximately 8 billion linearly independent statistics (including the 1.5 billion statistics in the redistricting data), but the demographic and housing characteristics NMF contains 25 trillion linearly independent statistics. It's so large that all the storage space on the Census Bureau's data dissemination server farms couldn't hold it, even compressed. 

The NMFs were developed in a modern, massively distributed cloud computing environment. That's the only environment where it makes sense to analyze them. The NMFs are experimental, and the suggestions in \citet[MSI, hereafter]{mccartan:etal:2023} for improving the documentation are valuable contributions. The current documentation was developed in just a few weeks by the same staff working on the official products, which were appropriately prioritized. As experimental products, the expectation is that users will provide the kind of feedback that MSI have provided, and that this feedback may be reflected in future products. 

The NMFs were not designed for direct publication. The internal NMFs used in the production DAS could not be released because their storage format commingled confidential information with the noisy measurements. New software was written to extract the noisy measurements, embedded in a 400TiB pickled data structure, and array them in the 40TiB Parquet files used for publication. This is why the noisy measurements could not be released under a Freedom of Information Act (FOIA) request, as the Census Bureau correctly responded when it received and denied such a request on December 1, 2022 \citep[p. 2]{22-cv-09304-JSR:dismissal}. Justin H. Phillips, represented by the Election Law Clinic at Harvard University, then sued the Census Bureau to reverse the FOIA decision. The plaintiff continued to press the suit even after the agency agreed to publish the noisy measurements after the reformatting described above as an experimental data product with documentation. Neither the reprogramming nor the documentation would have been required in a FOIA release. But the plaintiff would not agree to dismiss the suit until the Census Bureau agreed to a firm publication date for the Redistricting Noisy Measurement File of August 23, 2023 \citep[p. 4]{22-cv-09304-JSR:dismissal}. The Assistant U.S. Attorney for the Southern District of New York and the plaintiff agreed to dismiss the suit with prejudice given that publication deadline \citep[p. 4]{22-cv-09304-JSR:dismissal}. This is exactly what happened: the NMFs were approved for release, as the original scholars had requested, and sufficient documentation was developed to permit their use without delaying any other 2020 Census data products. The plaintiff pressed for attorneys' fees arguing that the suit had ``substantially prevailed'' in forcing the Census Bureau to release the NMFs. The court was sympathetic to this argument but ultimately held that ``the Census Bureau has provided persuasive evidence that it was intending to release files containing the noisy measurements data, like those requested by plaintiff, well in advance of plaintiff filing his FOIA request and lawsuit'' and denied attorneys' fees \citep[p. 10]{22-cv-09304-JSR:opinion}.

As MSI note, the experimental NMFs cannot be used without a working understanding of the production DAS code base, which is also public \citep{USCensus:2021:productioncode}. They note, in particular, three improvements that could be made to the existing noisy measurement files:  (1) centralized documentation and codebooks to properly read and format the file, (2) aggregation specifications that link queries to tabulated statistics, and (3) full block assignment files for NMF geographies. MSI make other suggestions that appear to require major programming investments:  (1) unnesting the NMF Parquet files, (2) filling in missing geographies that are on or off the 2020 DAS spine, and (3) providing application programming interface access to the full set of NMFs. Given that the 2020 NMFs are an experimental data product whose improvements do not directly support the 2030 Census, the most productive next step for improving the quality of the public NMFs would be for a group of scholars like MSI to secure a National Science Foundation, National Institutes of Health, Sloan Foundation, or other grant to develop tools to analyze the NMFs. That grant  should include adequate budget for Census Bureau staff time, expert consultants, and cloud computing resources. With these resources in hand, in particular direct Census Bureau staff salary commitments, it would be reasonable for the scholars to set the priorities for improving the usefulness of the existing 2020 Census NMFs. Absent such an intellectual and financial commitment, it is hard to find fault with the Census Bureau's prioritization of 2030 Census research.

Excellent models for how dedicated scholars like MSI can collaborate with the Census Bureau already exist. These include the Longitudinal Employer-Household Dynamics Program \citep{abowd:etal:2009}, Business Dynamics Statistics \citep{haltiwanger:etal:2013}, Opportunity Atlas \citep{chetty:etal:2020}, Criminal Justice Administrative Record System \citep{finlay:etal:2022}, and Census-Enhanced Health and Retirement Study \citep{isr:undated}. The Census Bureau cannot be expected to provide all the resources to make an experimental product better. Productive engagement means that external scholars commit to sharing the development costs and to putting the resulting enhanced product and documentation in the public domain.

There are now official publications from the 2020 Census that do release the noisy measurements and the margins of error associated with the noise added by the differential privacy mechanism, for example, the 2020 Census Detailed Demographic and Housing Characteristics File A (Detailed DHC-A). This data release demonstrates that when the disclosure avoidance system is designed from the outset to publish the noisy measurements, a statistically better product can be released. The Detailed DHC-A uses the same differential privacy framework as the DAS ($\rho$-zero-concentrated differential privacy) and the same mechanism (discrete Gaussian noise), but the implementation has, approximately, one noisy measurement for each publication statistic. Design constraints imposed on the DAS that caused the redistricting and DHC data to have massively more noisy measurements than publication statistics (such as the requirement to generate synthetic microdata, which required noisy measurements to estimate all possible interactions between variables) were relaxed for the Detailed DHC-A, permitting publication of the noisy measurements with only very minor post-processing. Virtually every one of the 500 million statistics in the Detailed DHC-A is the noisy measurement itself.

What I want to do in the rest of this commentary is focus researchers on the real issue: Why did the Census Bureau design the DAS to transform noisy measurements via post-processing into microdata to publish conventional tabular summaries? My hope is that researchers can suggest acceptable modifications to the official products that still meet their dominant use cases while better supporting the kinds of statistical inferences that MSI suggest and still properly protecting confidentiality. I focus on the redistricting use case because it drove virtually all the design decision-making for the first application of the 2020 DAS, which was then extended and enhanced for its second application to the DHC. The discussion is not a rehash of old arguments. It is a focused attempt to explain to nontechnical readers why the redistricting data took the form they have and what, if anything, the Census Bureau and redistricting community can do to improve the published data product in future censuses. To serve that goal, the reader must understand what determined the publication constraints and what options future censuses might employ. Even more importantly, readers must understand that over the course of the inter-censal decade, all parties to the production of redistricting data must consider and ultimately agree to changes in their content and format.

\section{Thinking about Redistricting Noisy Measurements from First Principles}
\label{sec:first_principles}
Using the differential privacy framework, the engineering begins with the \emph{query workload}, which is the technical term for the collection of statistics for which the data steward wishes to control the error in anticipation of publication. For the 2020 redistricting data, the query workload is a set of tables that display 252 counts for the resident population, 8 counts for the group quarters population, and 2 counts for the housing units---a total of 262 statistics---for 8.6 million geographies composed of aggregations of a prespecified atom, the census block, that tessellates the United States---divides its physical area into mutually exclusive and exhaustive pieces. This query workload is the result of a decade-long negotiation between the Census Bureau's Redistricting and Voting Rights Data Office---part of the Decennial Census Programs Directorate---and the National Conference of State Legislatures---a nonpartisan body encompassing representatives from all 50 states and the District of Columbia---that culminated, as in previous decades, with the Federal Register Notice in 2018 \citep{2018:FRN:PL91171} containing the detailed specifications for these statistics. Following that publication, state, county, municipal, tribal, and other organizations empowered their own redistricting offices to develop software that ingested the official redistricting data, combined them with other data, and produced tentative new voting districts. This software runs immediately after the release of the official data to produce legislative districts whose boundaries are defined using the same census blocks as the official redistricting data. A considerable investment, also occurring over the full decade, ensures that the boundaries embodied in the geographic areas from the nation to the census block properly reflect the boundaries of the political entities whose legislative bodies require new districts.

The second element in engineering a typical differentially private data publication system is the \emph{query strategy}, which is the technical term for the collection of statistics to which the privacy-loss budget will be allocated. The statistics in the query strategy are calculated directly from the confidential microdata. Then, they are passed to the chosen differential privacy mechanism along with the privacy-loss budget allocation assigned to each statistic. For each statistic, the mechanism draws random bits, transforms them as specified to achieve the appropriate probability mass function, and adds that random number to the confidential statistic. The output of the mechanism is called a \emph{noisy measurement}. The collection of noisy measurements from the 2020 Census redistricting data, referenced by MSI, is called the 2020 Census Redistricting Noisy Measurement File.

If the query strategy for the 2020 redistricting data were identical to the query workload, then there would be exactly one noisy measurement for every linearly independent statistic in the published data. Because the differential privacy mechanism in the DAS used discrete Gaussian noise, the variance of each published statistic would, in this case, depend only on the privacy-loss budget allocated to that statistic and could be easily calculated from the formulas in \citet{canonne:etal:2020}. Because the DAS added independent discrete Gaussian noise to each confidential statistic, the variance of any aggregation of the 2020 redistricting data could, again in this case, be calculated by summing the variances of each independent statistic composing the aggregation. To be succinct, for any proposed new voting district, the variance of its total population and of the component racial and ethnic subpopulations would, in this case, be the sums of the variances of those statistics in the geographic areas aggregated to produce the district.\footnote{Some subtlety is required here, as noted in MSI. To calculate the minimum variance version of the desired statistic when aggregating over geographic areas, the user of the NMF must compute the minimum off-spine distance geographic aggregation. The formula is in \citet{cumingsmenon:etal:2023}, and the code is in the DAS production code base \citep{USCensus:2021:productioncode}. A stand-alone Census Bureau-supported version of this calculation would be a valuable addition to the NMF toolkit.} 

Table 7 in \citet{abowd:etal:2022} reveals that the query strategy for the redistricting data was much more extensive than the query workload. The query workload consists of the 1.5 billion linearly independent statistics noted in Section \ref{sec:origins}. The query strategy consists of 16 billion linearly independent statistics---meaning that the NMF is an order of magnitude larger than the 2020 redistricting data. Why did this happen? Should it be repeated? Can it be improved? These are the research questions that experimental products like the NMF support.

\section{Why Is the Query Strategy So Much Larger than the Query Workload in the 2020 Redistricting Data?}
\label{sec:why}

The basic reason that the redistricting data query strategy produced 16 billion statistics instead of 1.5 billion is the design constraints imposed by consensus among the Census Bureau's career senior executives---the members of its Operating and Data Stewardship Executive Policy Committees, including me. These constraints were that the 2020 Disclosure Avoidance System:

\begin{enumerate}
    \item could not affect the apportionment use case;
    \item had to use the query workload already defined by the 2018 Federal Register Notice for the 2020 Redistricting Data (P.L. 94-171) Summary File;
    \item had to accept the 2020 Census Edited File (CEF) as input;
    \item had to deliver its output to the 2020 Census tabulation system as unweighted microdata respecting the schema used for the CEF;
    \item could not delay the publication of the 2020 Redistricting Data (P.L. 94-171) Summary File beyond March 31, 2021.
\end{enumerate}

To meet the first constraint, the total populations of the 50 states and the District of Columbia, as well as the population of the Commonwealth of Puerto Rico, were not perturbed. The second constraint meant that a value for every one of the 1.5 billion statistics specified in the workload had to be published. This constraint effectively ruled out adaptive implementations, such as the one used for the recently released Detailed DHC-A, that could have tested thresholds, then aggregated cells according to specified precision (inverse variance) targets. The third constraint meant that the definition of the confidential data was the record-level image in the CEF and not the actual response data, which were contained in the 2020 Census Unedited File (CUF). The fourth constraint meant that the noisy measurements had to be post-processed into microdata forcing the estimation of many statistics not in the query workload. The fifth constraint, which was unofficially relaxed to August 12, 2021 due to the pandemic, meant that the processing required to run the DAS could not push the publication date beyond the statutory deadline. Programming efficiencies produced code that ran in less than a day for the redistricting data, allowing adequate time for the scheduled human reviews of the output.

2020 DAS post-processing of the noisy measurements into microdata had two important mathematical consequences. First, it implied non-negativity constraints on all cells. Second, it implied that some privacy-loss budget had to be allocated to what is called the \emph{detailed query}: the interaction of all the tabulation variables with each other. For the redistricting data, the person-level detailed query has 2,016 cells---far more than the 260 person-level cells in the published tables. Every one of these detailed query cells, as Table 7 in \citet{abowd:etal:2022} shows, got the lion's share of the privacy-loss budget in the 2020 DAS---far more than the core query that represented the 252 cell workload of $VOTINGAGE \times HISPANIC \times CENRACE$ (notation from \cite{abowd:etal:2022}). 

As \citet{abowd:etal:2021} showed, the non-negativity constraints are the accuracy culprit. They create a post-processing error that cannot be reasonably controlled via the privacy-loss budget; however, they can be greatly reduced by algorithmic tuning, which was accomplished for the redistricting data. There are good reasons to do this post-processing in spite of the error because the post-processing discovers sparsity in the confidential tables, which permits most of the zeros in the confidential data to appear as zeros in the published data. Evaluating the post-processing error has nothing to do with the values of the noisy measurements, and can only be studied in combination with the published tables and assessment of the simulation properties of the 2020 DAS, which a team at the Census Bureau is doing (e.g., \citet{cumings-menon:2024}). 

The 2020 Census is not the only recent population census to encounter this issue. The publication tables from the 2021 Census of England and Wales published by the Office of National Statistics (ONS) in the United Kingdom have exactly the same error due to non-negativity constraints even though the noise added to their publication tables was not produced using a differential privacy framework. When adding the noise, ONS processed the tables to eliminate negative values and constrain population margins to sum to fixed population totals. This processing, which is lighter-touch than the DAS post-processing required to produce microdata, nevertheless means that in 2021 Census of England and Wales cells with low counts have a slight positive bias and cells with high counts a slight negative bias \citep{ONS:2023}.

The privacy-loss budget allocations shown in Table 7 of \citet{abowd:etal:2022} were determined by extensive experiments documented therein. Those experiments were necessary because the atom of the geographical hierarchy in the redistricting use case cannot have a minimum population, and the tables at every geographic level of the redistricting data must have an entry in every cell. Because of these design constraints, there can be neither minimum populations in a census block nor block-level tables that aggregate race and ethnicity combinations. Thus, the work-horse geographic unit became the census block group, but not the publication block group---rather a custom block group that isolated blocks containing group quarters or tribal areas from the rest of the blocks in the block group \citep{cumingsmenon:etal:2023}. Block groups have large enough populations to proxy for a voting district in lower-population municipalities like census places. The work-horse table became the detailed query because it prevented the group quarters population characteristics from cross-contaminating the housing unit population characteristics when the microdata were created. Neither of these outcomes is essential to the confidentiality protection framework. They are consequences of the design constraints listed above, particularly imposing non-negativity on every cell, implementing hierarchical consistency via tabulation from microdata, and forcing every cell in every table to have a value (no collapsing of cells). As noted in related work \citep{kenny:etal:2024} ``the NMF contains too much noise to be directly useful without measurement error modeling, ... TopDown's post-processing reduces the NMF noise and produces data whose accuracy is similar to that of swapping.'' That is precisely point of the post-processing, and confirms the analysis in \citet{wright:irimata:2021}. If the redistricting community, especially the academic contributors, want the noisy measurements to be less noisy, some of the constraints must be relaxed. Relaxing those constraints requires consensus in the redistricting community, practitioners as well as academics, concerning the geographic atom and the publication format.

\section{How Might the Official Products Be Different?}

The redistricting community largely drove the content and form of the 2020 Census redistricting data. Unless the Census Act is amended or the Voting Rights Act is repealed or further weakened by the Supreme Court this decade, the redistricting community will drive the content and form of 2030 Census redistricting data. That is how the statutory mandate to produce these data works. Now that the Census Bureau has acknowledged that publication formats that include detailed tables for geographies like census blocks require much more comprehensive confidentiality protections, there's no going back to the systems used for the 2010, 2000, and 1990 Censuses. 

Redistricting researchers could ask: Given a fixed privacy-loss budget, what format for the publication tables best meets the redistricting use case? That format could be the one embodied in the 2020 DAS, which does an excellent job of preserving zeros in the confidential data (more than 90\% of the confidential zeros are zeros in the published tables\footnote{This statement is based on an assessment of the final 2010 demonstration redistricting data generated by the DAS as compared to the originally published 2010 redistricting data. No access to confidential data is required to confirm it.}) while still carefully protecting confidentiality. Or, it could be the application of the full privacy-loss budget to the publication query (the 262 cells in the 2020 redistricting data tables) with minimal post-processing. Those tables would have many negative entries, but every entry would be unbiased (at least from the disclosure avoidance noise), and their aggregates might produce low-variance voting district statistics that could withstand Voting Rights Act scrutiny. Or something in between. Or something completely different.

\subsection*{Disclosure Statement}
The author has no conflicts of interest to declare.

\subsection*{Acknowledgments}
The author is the former chief scientist and associate director for research and methodology at the U.S. Census Bureau. The opinions expressed in this commentary are his and not those of the Census Bureau. He acknowledges thousands of insights gained from years of working with the teams at the Census Bureau, both internal and external, who implemented and assessed the 2020 Census Disclosure Avoidance System. Dan Kifer and Philip Leclerc provided helpful comments on this article.
 
\subsection*{Contributions}
JMA researched and wrote this article.

% All references should be stored in the file "references.bib"
% Please do not modify anything below this line.
\printbibliography

\end{document}